# Analysis of the Annealing Budget of Metal Oxide Thin-Film Transistors Prepared by an Aqueous Blade-Coating Process

*Tianyu Tang, Preetam Dacha, Katherina Haase, Joshua Kreß, Christian Hänisch, Jonathan Perez, Yulia Krupskaya, Alexander Tahn, Darius Pohl, Sebastian Schneider, Felix Talnack, , Mike Hambsch, Sebastian Reineke, Yana Vaynzof, Stefan C. B. Mannsfeld \**

T. Tang, P. Dacha, J. Kress, F. Talnack, Dr. K. Haase, Dr. M. Hambsch, Prof. Y. Vaynzof, Prof. S. C.B. Mannsfeld
Center for Advancing Electronics Dresden (CFAED)
Faculty of Electrical and Computer Engineering
Technische Universität Dresden
Dresden 01069, Germany
Email: tianyu.tang@tu-dresden.de, stefan.mannsfeld@tu-dresden.de

J. Kreß, C. Hänisch, Prof. S. Reineke, Prof. Y. Vaynzof
Dresden Integrated Center for Applied Physics and Photonic Materials (IAPP)
Institute for Applied Physics
Technische Universität Dresden
Dresden 01069, Germany

J. Perez, Y. Krupskaya
Leibniz Institute for Solid State and Materials Research (IFW)
Dresden 01069, Germany

A. Tahn, S. Schneider, Dr. D. Pohl
Dresden Center for Nanoanalysis (DCN)
Center for Advancing Electronics Dresden (CFAED)
Technische Universität Dresden
Dresden 01069, Germany

Keywords

**Abstract**

Metal oxide (MO) semiconductors are widely used in electronic devices due to their high optical transmittance and promising electrical performance. This work describes the advancement toward an eco-friendly, streamlined method for preparing thin-film transistors (TFTs) via a pure water-solution blade-coating process with focus on a low thermal budget. Low temperature and rapid annealing of triple-coated indium oxide thin-film transistors (3C-TFTs) and indium oxide/zinc oxide/indium oxide thin-film transistors (IZI-TFTs) on a 300 nm $SiO_2$ gate dielectric at 300 °C for only 60 s yields devices with an average field effect mobility of 10.7 and 13.8 $cm^2\ V^{-1}\ s^{-1}$, respectively. The devices show an excellent on/off ratio (>$10^6$), and a threshold voltage close to 0 V when measured in air. Flexible MO-TFTs on polyimide substrates with $AlO_x$ dielectrics fabricated by rapid annealing treatment can achieve a remarkable mobility of over 10 $cm^2\ V^{-1}\ s^{-1}$ at low operating voltage. When using a longer post-coating annealing period of 20 min, high-performance 3C-TFTs (over 18 $cm^2\ V^{-1}\ s^{-1}$) and IZI-TFTs (over 38 $cm^2\ V^{-1}\ s^{-1}$) using MO semiconductor layers annealed at 300 °C are achieved.

## 1. Introduction

Post-transition metal oxide thin-film transistors (MO-TFTs) moved into the focus of current research because of high carrier mobility, optical transparency, flexibility, and large-area uniformity.[1, 2] Solution-processed metal oxide (MO) semiconducting films are expected to enable a low-cost, fast emerging technology to revolutionize large-area electronics.[3-7] The excellent electrical performance of indium oxide ($InO_x$) makes it an attractive semiconductor material since the extensive 5s orbital overlap leads to a broad conduction band.[8, 9] However, the stability of solution processed MO semiconductors is often lowered by environmental moisture/water, oxygen and etchants that is problematic for many low power electronics applications.[10, 11] When fabricated at temperature <=300 °C with fast annealing, MO TFTs enjoy performance benefits over n-type organic semiconductors and yield thin, dense films at low cost.[12] The simplicity by which such films can be fabricated and the high quality of the resulting devices have already led to commercial device fabrication with only minor changes to the fabrication scheme.[13]

Various methods have been proposed to improve the performance of MO films by increasing the metal-oxygen-metal (M-O-M) network content[14] and by controlling the oxygen vacancies. Promising techniques implemented include using organic-based (OB) solvents to trigger the condensation reaction at low temperatures (e.g., 2-methoxyethanol,[15] dimethylformamide[16]). Additional combustion fuels (e.g., acetyl acetone, benzoyl acetone, and 1,1,1 Trifluoroacetylacetone)[17-19] were used to reduce the oxide conversion temperature. These techniques enabled filling in gap trap states to improve electron transport in inorganic electron donors (Li, Zn, Ga, Sn, etc.),[20-23] by organic bulk doping to intentionally frustrate the crystallization.[24-26] Nevertheless, most of the previous reports on solution-processed MO films utilize toxic solvents, which result as a potential threat to the environment. Liu et al. reported the MO-TFTs by using an aqueous precursor solution and studied the annealing time and low temperature effect.[27-29] They observed that a lower annealing temperature with sufficient annealing time results in dihydroxylation and condensation process in such a system. However, unless they are annealed for long times, MO-TFTs based on aqueous precursor solutions still exhibit low mobilities—a problem that still needs to be addressed.

Other successful strategies implemented for the MO TFTs performance enhancement are channel structure engineering including self-passivated multilayers,[30] organic-MO based TFTs,[31] homojunction[32] and heterojunction[33, 34] structures, and high-k dielectric utility.[35, 36] Usually, deposition of the MO semiconductor films using a layer-by-layer (LBL) process are preferred to fill in cracks and pores, leading to smoother, denser films. However, this process is laborious and difficult to reconcile with the concept of an efficient and continuous process in industrial production. Another issue with MO-TFT is the on-off ratio that deteriorates to an "always-on" state when the semiconductor thickness reaches a certain thickness due to the large number of free carriers.[37] If MO-TFTs are fabricated in ambient atmosphere, it was found that fabricating them in steps—layer by layer—increases air stability because it helps protecting against post-fabrication chemical degradation that would reduce the electron mobility.[10, 38]

There are many reports in which solution processing is explored to prepare 2D MO semiconductor films on large-scale substrates with simple technologies at low annealing temperatures (200–300 °C).[39, 40] In addition, the device layouts and layer structures are being engineered for highest possible performance. Anthopoulos et al. confined the free electrons at the critical heterojunction interfaces to dramatically enhanced electron mobility up to 45 cm$^2$ V$^{-1}$ s$^{-1}$ by structuring different MO films in 2D electron gas (2DEG) systems.[41, 42] Facchetti and coworkers structured InO$_x$ films by polymer doping to achieve a similar effect, opening up a new pathway to organic–inorganic bulk hybrid devices.[43] Heterojunctions provide guidelines for the design of MO-TFTs with performance characteristics potentially superior to most MO semiconductor technology.

Recently, research has largely focused on InO$_x$ and ZnO$_x$ based devices due to their high mobility, scalability and low temperature processing.[44, 45] Among the extensive solution-processed manufacturing methods (spin-coating, blade-coating, spray-coating, inkjet-printing, etc.)[46, 47] blade coating is considered not only a promising method for ultra-thin MO semiconductor films with high stability, but also has been demonstrated to reduce film-processing time and improve industrial compatibility with the films having strong crystallographic texture, planar surface topography and strong inter-grain connectivity.[48-50] However, there are as of yet no studies that show the evolution of metal oxide formation during the fabrication process of the MO films, especially the initial transformation process of the precursor material to MO films and how this is linked to the evolution of the device electrical performance.

In this study, we demonstrate the fabrication of water-based solution processed ultrathin MO film via blade coating at low temperature of 300 °C with a focus on reducing the thermal budget of this post-coating annealing process as much as possible while still only using pure DI water as a solvent without any additives. For this purpose, we studied films fabricated from a single coating step with different annealing times, ranging from few seconds to 20 min. Ellipsometry and TEM were used to obtain a quantitative understanding of $InO_x$, $ZnO_x$, and $AlO_x$ film thickness evolution for different annealing times. In order to link the observed thickness variations to any possible change in the chemical composition of the films, X-ray photoelectron spectra (XPS) were recorded for samples of the different processing stages with respect to the minimum annealing time required to produce MO from the precursors. Ultraviolet photoelectron spectra (UPS) are used to discuss the heterojunction engineering with different oxides to improve the performance of TFTs.

We also characterized the electron field-effect mobility $\mu_e$ of MO-TFTs in air. Unlike the films used in the morphological analysis of the annealing time impact, these devices were fabricated using three coating steps to maximize performance. The mobility of heterojunction $InO_x/ZnO_x/InO_x$ thin-film transistors (IZI-TFTs, 13.8 cm$^2$ V$^{-1}$ s$^{-1}$) is found to be consistently superior to the triple-coated indium oxide thin-film transistor (3C-TFTs, 10.7 cm$^2$ V$^{-1}$ s$^{-1}$) (MO films prepared at 300 °C for 60 s). Remarkably, we demonstrate IZI-TFTs heterojunction transistors based on $AlO_x$ dielectic on polyimide films with an average electron mobility of 10.3 cm$^2$ V$^{-1}$ s$^{-1}$ with semiconductor and dielectric films both annealed at 300 °C and for only 60 s. In tests of the gate bias stress reliability, we measured at least five TFTs from different samples for each TFT structure that show nearly indistinguishable results. The excellent operating performance and stability of our devices, coupled with the eco-friendly fabrication process indicates that the manufacturing of transparent circuits for large area device architectures is widely possible.

**2 Result and Discussion**

## 2.1 Blade-Coating and Thin-Film Transistor Fabrication Process

The precursor solutions as prepared were deposited using a blade coating method called "solution shearing" with a silicon oxide blade at an optimized substrate temperature of 80 °C. As the meniscus front moved with the blade, a uniform thin film was deposited.

The hard bake of these films was conducted by thermal annealing at 300 °C in air at varied time (0–1200 s.) to obtain the desired MO film. The individual layers in the multi-layer devices were each prepared in identical fashion (**Figure** 1c,d). The combustion process that occurs during the annealing densifies the MO precursor film and leads to the formation of an M-O-M framework.[19, 51] Finally, aluminum (Al) source and drain (S/D) electrodes were deposited by thermal evaporation to act as the top contacts as shown in Figure 1e. The channel width and length for all devices are 1000 and 100 µm, respectively. The process detail and the schematic evolution process of MO-precursor coordination chemistry are illustrated in Figure 1.

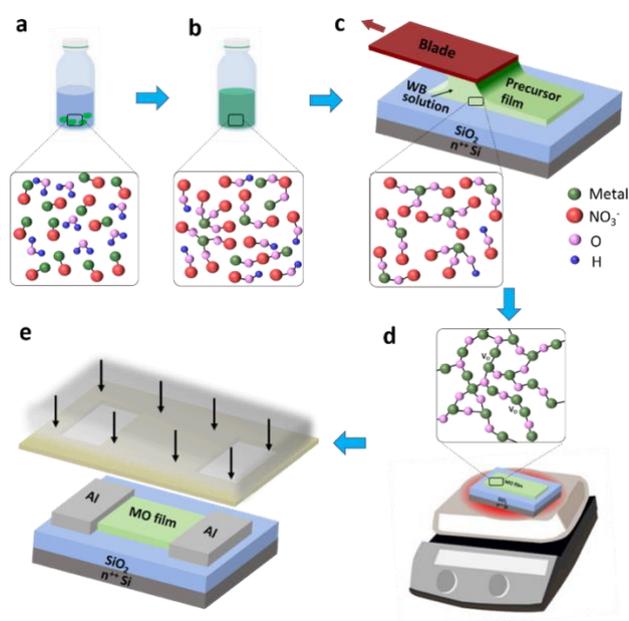

**Figure 1.** Schematic of device fabrication and evolution of MO precursor coordination chemistry. a) In/Zn nitrate salts dissolved in deionized water. b) The MO precursor solution obtained by stirring and aging. c) Representation of the involved blade coating process. d) Thermal annealing process to produce dense MO films by combustion synthesis e) bottom gate, top-contact transistor fabricated with Al evaporation through a shadow mask.

## 2.2. Chemical structural characterization of MO films

In order to probe for any potential changes in the composition of the MO films with annealing time, we performed X-ray photoemission spectroscopy (XPS) analysis on MO films that were annealed at 300 °C for different bake times. **Figure** 2a displays the O1s spectra measured by XPS for InO$_x$ and ZnO$_x$ precursor films as they evolve during annealing at 300 °C for 10, 15, 30, 60, 120, 300, 600, 900, and 1200 s. In case of InO$_x$, the O1s peaks are deconvoluted into three main peaks that are located at 529.7 ± 0.2, 531.5 ± 0.2, and 533.2 ± 0.2 eV. According to literatures,[24, 52] the peak at 529.7 eV is ascribed to the O$^{2-}$ ions associated with the neighboring metal ions of original metal-oxygen-metal (In-O-In) from the In$_2$O$_3$ lattice. The peak at 531.5 eV is related to the oxygen-deficient region (In-O$_{vac}$), which is caused by nonstoichiometric defects in In$_2$O$_3$ while the peak at 533.2 eV is from the adsorbed oxygen or weakly bound species hydroxyl groups (In-OH). The relative contributions of these three species to the O1s signal are summarized in Table S1(Supporting Information). The peak corresponding to In-O-In remained largely unchanged at ≈36.5% when the films were subject to annealing from 15 to 120 s. However, when the annealing time exceeded 300 s, the percentage of In-O-In increased from 37.2% (300 s) to 40.0% (600 s). Interestingly, in the same time intervals, the relative quantity of In-OH increased from 54.4% (10 s) to 61.7% (120 s), and then slightly decreased from 62.1% (300 s) to 59.5% (600 s). The relative quantities of In-OR progressively decreased during annealing from 8.6% at 10 s to 0.2% at 1200 s. Taken together these results indicate that the crystalline quality is ameliorated through the annealing process.

A similar analysis of the O1s peak measured on ZnO$_x$ films revealed a strong peak at 530.5 ± 0.2 eV and two distinct shoulders at 532.2 ± 0.2 and 532.8 ± 0.2 eV.[52, 53] The peak at 530.5 ± 0.2 eV is associated with the Zn☐O☐Zn from ZnO lattice. The peak at 532.2 can be attributed to the oxygen vacancies (Zn-O$_{vac}$).[54] The peak at 532.8 ± 0.2 eV could be assigned to Zn☐OH due to the presence of hydroxyl groups. The relative contribution of the three species is summarized in Table S2 (Supporting Information). Upon annealing, the Zn☐O☐Zn contribution increases slightly, from 9.7% at 10 s to 10.8% at 1200 s. This increase is accompanied by a significant increase in the Zn☐O$_{vac}$ signal from 45.3% to 59.4% in the same time period. On the other hand, the Zn☐OH species decreases from 45.3% to 29.8%. These results suggest that the hydroxide groups are mainly converted to Zn☐O$_{vac}$ during the annealing process.

**Figure 2**. a) XPS O1s spectra and b) optical band gap of InO$_x$ and ZnO$_x$ film annealed for different times at 300 °C in air. c) Band diagram of InO$_x$ and ZnO$_x$ annealed for 10 and 1200 s. d) XPS survey spectrum showing the characteristic peaks for In, Zn, O, and C in InO$_x$ and ZnO$_x$ films annealed at 300 °C for 60 s. Inset image shows the In and Zn core level peaks.

The chemical conversion of the precursor material to form MO semiconductor films with annealing process has been investigated.[55-57] The decomposition of nitrate salts during the annealing process is consistent with the formation of oxides of nitrogen through some intermediate products can be written as follows:

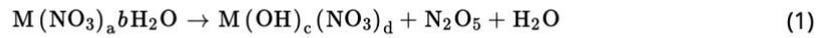

$$M(NO_3)_a bH_2O \rightarrow M(OH)_c(NO_3)_d + N_2O_5 + H_2O \quad (1)$$

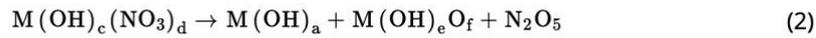

$$M(OH)_c(NO_3)_d \rightarrow M(OH)_a + M(OH)_e O_f + N_2O_5 \quad (2)$$

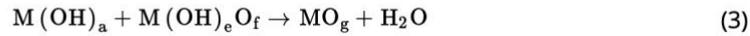

$$M(OH)_a + M(OH)_e O_f \rightarrow MO_g + H_2O \quad (3)$$

As shown in Equations 1-3, the MO semiconductor materials can be produced by poly-condensation and dihydroxylation/dehydration process during the annealing process (in Figure 1). The precursor films were dried on the hotplate at a temperature of 80 °C after the deposition by the blade-coater. However, at this stage, the precursor films still contain significant amounts of H$_2$O. After this initial treatment the films are annealed at 300 °C. Within the first 10 s, most of the solvent H$_2$O is vaporized and removed from the films. In fact, a low content of M-O-M framework in the MO films has been formed in the first 10 s (Table S1, Supporting Information). The O 1s peak indicates that long annealing treatment further lowers the contributions of the metal-hydroxyl (M-OH) species and increases the metal oxide bond content formation to form M-O-M species. Secondly, the hydroxyl groups M-OH are dehydroxylated and dehydrated during the annealing treatment under release of H$_2$O, and this process mainly occurs between 10 and 30 s annealing time. Usually, the reaction process that produces oxygen vacancies is slow, having been estimated to tens of minutes in literature reports.[58, 59] By the end of the first minute (60 s), most of the hydroxyl groups have reacted and oxygen vacancies have begun to get filled by free oxygen atoms. During the next 4 min (until 300 s), the oxygen atoms continue to fill the existing oxygen vacancies and convert them into M-O-M groups. In general, with prolongation of the annealing treatment time, the MO films is mainly composed of M-O-M species, whereas for short annealing treatment, the precursor decomposition is not complete and there would be surplus M-OH species from the precursor solution.

The work function and valence band position of $InO_x$ and $ZnO_x$ films annealed at 300 °C for different times were measured by UPS. All the thin MO films were deposited on conductive heavily doped $Si^{++}$ substrates. The measurements were performed with a negative bias of − 5 V to ensure no charging would occur in the samples. Here, He I (21.22 eV) was utilized as a photon source for the UPS measurement.

The UPS data from the $InO_x$ films annealed at 300 °C for different times show the work function ($E_\Psi$) decreases with longer annealing times, progressively changing from 4.62 eV at 10 s to 4.20 eV at 1200 s. A similar trend is observed for the valence band minimum ($E_{VBM}$) from $InO_x$ films, decreasing from 7.62 eV at 10 s to 7.22 eV at 1200 s. (Figure S1, Supporting Information). Likewise, the $ZnO_x$ films showed a similar decrease in $E_\Psi$ from 4.06 eV at 10 s to 3.86 eV at 1200 s and $E_{VBM}$ from 7.36 eV (10 s) to 7.23 eV (1200 s) (Figure S2, Supporting Information). As the annealing time is increased further, the Fermi levels of $InO_x$ and $ZnO_x$ get closer to the conduction band edge $E_{CBM}$, which has been reported to improve the n-type TFT performance.[60] The respective band gap diagrams of $InO_x$ and $ZnO_x$ annealed with different times are estimated in Figure S3 (Supporting Information).

As can be seen from Figure 2b and Figure S3 (Supporting Information), for the as-deposited (only annealed on the hotplate at 80 °C) film and the film annealed for 10 s, the $E_G$ of the $InO_x$ and $ZnO_x$ films are 3.79 and 3.59 eV, respectively. As the annealing time increases from 10 to 30 s, the optical band gap decreases from 3.79 to 3.73 eV for $InO_x$ (3.59 to 3.57 eV for $ZnO_x$), which is attributed to more M-O-M bonds being formed as a result of the M-OH group dehydroxylation and dehydration. After an annealing time of 300 s, the InO$_x$ and ZnO$_x$ films have the smallest band gap of 3.64 and 3.48 eV, respectively. This has been attributed to an increased overlap in the In 5s atomic orbitals that itself results from the increased crystalline order in the film with increasing annealing time.[61] For annealing times between 300 and 1200 s, the optical band gap of InO$_x$ increases again from 3.48 to 3.70 eV, and the ZnO$_x$ band gap increases from 3.48 to 3.57 eV. This can be explained by the Brustein–Moss effect that describes an effective band gap widening phenomenon when the carrier concentration increases.[62, 63] This effect arises from the filling of trap states near the CBM in MO films. As is known, the oxygen vacancies are the source of carrier generation in n-type MO semiconductor. And an increase in free carriers could fill trap states that are associated with the MO thin film grain boundaries. Conversely, when these states are increasingly filled already, trapping of electrons moving the conduction band is reduced.[64] Since in our XPS experiments (O 1s spectra) we indeed saw such an increase in the oxygen vacancy concentration with increasing annealing time, we believe this effect explains the observed improved device characteristics in our MO films. In general, increased concentrations of oxygen vacancies concentration could not only elevate the Fermi level in MO films, but also shift $V_{TH}$ to the negative direction and raise the on-current to improve the field effect mobility in TFTs. The electrical properties will be discussed in Section 5.

Figure 2d shows the survey XPS spectra from 0 to 1200 eV that was corrected for specimen charging by referencing the C1s to 284.8 eV. In the inset of Figure 2c, the In 3d doublet spectrum is shown. For all of the samples the In 3d$_{5/2}$ and In 2d$_{3/2}$ are observed at 444.4 and 452.5 eV without significant peak shifts, which is indicative of In as trivalent In$^{3+}$ with those observed in In$_2$O$_3$ (Figure S4, Supporting Information). Similarly, in the inset of Figure 2d, the Zn 2p doublet appears at 1020.6 and 1044.5 eV, corresponding to Zn 2p$_{3/2}$ and Zn 2p$_{1/2}$ of all the samples, respectively (Figure S5, Supporting Information) corresponding to the state of Zn$^{2+}$ in ZnO. These observations are in good agreement with typical MO film values reported in previous reports.[65, 66]

## 2.3. Thickness and morphology characterization (Ellipsometry, TEM, and AFM)

The surface morphology is of utmost importance for device applications, since high surface roughness values can have a detrimental effect on device performance. To understand the MO-TFT performance variation, the morphology and microstructures of the MO films were characterized in detail. The film roughness is related to the film coalescence process. Higher annealing temperatures and longer annealing treatments could then increase the crystallinity, and the resulting larger grain size will affect the roughness of the films undoubtedly. Usually such observations are made for samples that are annealed for longer times and at higher temperatures than in this work. With the rapid annealing process at lower temperature studied here, we observed a slightly different variation of the roughness. The surface morphology of $InO_x$ and $ZnO_x$ films annealed at 300 °C for 10, 15, 30, 60, 120, 300, 600, 900, and 1200 s conducted by atomic force microscopy (AFM) are shown in **Figure** 3. Here, we observe that the $InO_x$ films exhibit a very low RMS roughness ($\sigma_{RMS}$) in the range of 0.15 (± 0.01) nm to 0.17 (± 0.01) nm. Similarly, $ZnO_x$ films show a similar surface roughness range of 0.16 (± 0.01) to 0.22 (± 0.02) nm with different annealing duration. Certainly, the initial change in $\sigma_{RMS}$ attributes to the dihydrogenation and dehydration process (10–60 s for $InO_x$ and 10–30 s for $ZnO_x$). In this process, water was vaporized and escape outside the precursor films at 300 °C, therefore the roughness of the film increases. As the crystallinity increases (60–300 s for $InO_x$ and $ZnO_x$), the films become dense and the roughness of the film decreases. Once the films are formed, the roughness remains almost constant under longer annealing treatment (300–1200 s for $InO_x$ and $ZnO_x$).

**Figure 3**. AFM images of a) $InO_x$ and b) $ZnO_x$ films annealed at 300 °C for different duration.

Based on initial experiments comparing optimized blade coating and optimized spin coating parameters, we believe that blade coating these films at a speed of 5000 µm s$^{-1}$ and a substrate temperature of 80 °C leads to a faster evaporation of the solvent that might explain the little-to-no variation in the $\sigma_{RMS}$ for the different annealing times. From this analysis, we observe that, independent of the annealing time, all films show exceptionally low surface roughness that is highly beneficial for device fabrication in general but also eases the fabrication of heterojunction devices in particular. The 3C heterostructure and IZI films showed similar values of $\sigma_{RMS}$ as the unilayer MO films.

In order to accurately assess the MO film thickness, we used ellipsometry measurements (Figure S7, Supporting Information). We found that the thickness of the MO film is significantly impacted by the chemical transformation during the annealing as shown in **Figure** 4. For $InO_x$ films, during the first 10 s, the remaining $H_2O$ in the film was evaporated rapidly and the thickness of the film drop dramatically from 5.7 to 4.3 nm. Annealing treatment for the next 60 s has a moderate effect on the film thickness, producing a further roughly 20% decrease from 4.3 to 3.4 nm, consistent with the dehydroxylation and dehydration of M-OH groups as observed from the UPS measurements during that annealing time window since the reduction in the films M-OH content and the formation of M-O-M are involving mass loss.

The HRTEM micrographs, obtained using the focused ion beam (FIB) technique, display a defined and uniform deposition of single-layer $InO_x$ film annealed at 300 °C for 10 and 300 s, respectively. Since we could not get a very clear interface image for the $InO_x$ film annealed for 10 s, we believe it to be still amorphous. However, the $InO_x$ film annealed for 300 s shows a clear lattice structure with a polycrystalline grain structure (Figure 4c).[67, 68] For the 300 s annealed sample we also used HRTEM to measure the film thickness that nicely confirmed the ellipsometry result of ≈3.7 nm. Figure 4c shows the $InO_x$ films exhibit micro-polycrystalline phase with obvious lattice structure visibility in the cross-sections.[67, 68] With longer annealing treatment, several well-defined Brag diffraction rings emerge as the MO films become polycrystalline. The strong peak is located at 2.22 $Å^{-1}$ corresponding to (222) plane of the cubic oxide phase in $In_2O_3$.[69] Additionally, the GIWAXS pattern of $ZnO_x$ films showed (100),

(002), and (101) diffraction rings related to a regular wurtzite crystal of ZnO, indicating the successful synthesis of ZnO in the ZnO$_x$ films (Figure S5, Supporting Information).[70] With the extension of annealing time after 600 s, the crystallinity of the MO films did not improve significantly. This indicates that the crystallization process is mainly completed before 600 s. Together these findings confirm the successful MO film formation from green solvent with a rapid annealing procedure.

The morphology evolution and the resulting film quality of multi-layer films and heterojunction MO films annealed at 300 °C for 60 s was also checked to confirm that there is no loss of thin film quality when going from a single coating step to triple-coated MO-TFTs. The corresponding HRTEM figures shown in Figure 4d,e readily confirm that the multilayer interfaces retain the low smoothness and high quality of the single coated layers. Figure 4ein particular, includes heterojunction cross-sections wherein the surface of InO$_x$, ZnO$_x$, and AlO$_x$ layer can be clearly distinguished. In agreement with the ellipsometry and HRTEM, the EDS mapping of the C, O, Al, In and Zn confirmed the existence of discrete nanometer-thin layers of InO$_x$ and ZnO$_x$ layers on a single coated AlO$_x$ (1C-AlO$_x$) film by an elemental sharp interface (Figure S9, Supporting Information) indicating the growth of ultra-thin InO$_x$(≈4.0 nm) and ZnO$_x$ (≈4.0 nm) layers on Si/SiO$_2$ substrates.

Similar to this work, Kirmani et al. also obtained ultrathin InO$_x$ films (as thin as 4 nm) be using a blade coating method.[71] Overall, we believe that the ability to produce ultrathin and smooth films are the main reasons why high quality M-O-M framework films can be obtained from additive-free DI water solutions that are annealed with a low thermal budget. We hypothesize that the efficient spread of the thermal energy in a thinner film with its very low film volume results in an equally rapid and efficient M-O-M formation and progression of the crystallization. Based on the results achieved here, we believe that our finding of needing only 60 s or less at 300 °C to produce high quality MO films despite the absence of any combustion promoters in our DI water precursor solution is due to achieving ultrathin films via blade coating.

**Figure 4**. Comparison of film thickness, RMS surface roughness for a) InOx and b) ZnOx films annealed at 300 °C for different times. c) HRTEM images of 1C-InOx film annealed at 300 °C for 10 and 300 s, d) 3C-InOx films, and e) IZI films on 1C-AlOx film stack cross-section annealed at 300 °C for 60 s.

**2.4. Thin-film Transistor Electrical Performance**

To shed light on the electrical properties of MO thin films, TFTs are fabricated in the common bottom-gate top-contact (BGTC) configuration with aluminum as source and drain metal contacts. All devices were fabricated and characterized under the same experimental condition in order to ensure a consistent comparison among all the devices characterizations.

2.4.1. InOx Multilayers TFTs

In this section we discuss the results for single-coated and multiple-coated InOx TFT devices. **Figure** 5ab, shows the blade coating equipment used to shear the films and the bottom-gate top-contact (BGTC) stack of the device for 3C-TFTs respectively. Figure 5d shows statistical results obtained from 30 TFTs with semiconductor channels of single-coated InOx TFTs (1C-TFTs), double-coated InOx TFTs (2C-TFTs), and triple-coated InOTFTs (3C-TFTs). The films are annealed in air for 60 s at 300 °C. The 1C-TFTs and 2C-TFTs exhibited a typical n-channel behavior with an average mobility of 0.6 cm$^2$ V$^{-1}$ s$^{-1}$ (average mobility from five working 1C-TFTs in all 30 TFTs) and 3.7 cm$^2$ V$^{-1}$ s$^{-1}$ (average mobility from 17 working 2C-TFTs in all 30 TFTs). Moreover, the 3C-TFTs showed an enhanced average mobility of up to 10.3 cm$^2$ V$^{-1}$ s$^{-1}$ (average mobility from 30 working 3C-TFTs in all 30 TFTs). It is fair to note that achieving mobility values of ≈10 cm$^2$ V$^{-1}$ s$^{-1}$ from only three coatings is a drastic improvement from prior attempts to rapidly annealed InOx TFTs where mobilities remained low (<1 cm$^2$ V$^{-1}$ s$^{-1}$).[29, 50]

**Figure 5.** Fabrication and electrical characterization of combustion-processed MO films. a) Photo of the blade-coating process; inset image shows the device size with Al electrodes with a scale bar of 1000 µm. b) Schematic of 3C-TFTs c) TFT transfer curves of single-coated, double-coated, and triple-coated InOx films annealed at 60 s at 300 °C. d) Mobility distribution of 30 devices annealed for 60 s at 300 °C. e) I–V curves of 3C-TFTs with each InOx films annealed for different time at 300 °C and f) 3C-TFT films (annealed at 300 °C for 60 s) on 300 nm Si/SiO2 substrate under low-voltage measurement.

In order to estimate the progression of performance with annealing time relative to the rapid (60 s) annealed films while still annealing only for an acceptably short period of time, we also built devices with 300 s annealed layers. Here, we studied triple-coated InO$_x$ TFTs (3C-TFTs) comparing them to 1C-TFTs and 2C-TFTs annealed for the same duration to get the sufficient understanding of the electrical performance evolution of the InO$_x$ films in relation to their thickness and number of coatings. Each shearing of a MO layer followed by a bake of 300 °C for 300 s in air. The measured transfer and output characteristics for increasing number of films and their corresponding evaluation of the critical device performance parameters are depicted in Figure 5c,d The resulting average mobility of 3C-TFTs of 15.8 cm$^2$ V$^{-1}$ s$^{-1}$ (all of 30 TFTs working) is further improved from the 3C-TFTs annealed for 60 s (10.7 cm$^2$ V$^{-1}$ s$^{-1}$).

The thickness of the MO semiconductor film also plays a major role in the electrical performance, which has been also reported by others.[10, 72] The increase in mobility with increasing InO$_x$ film thickness (from multiple coatings) is commonly attributed to a reduction in the number of trap sites in the InO$_x$ films.[73] It is noted that even defects at the top surface can modulate the channel performance in the typically very thin film thickness InO$_x$-TFTs.[71] Kim et al. propose the improved mobility with thicker InO$_x$ films is due to the passivation of top surface defects.[10, 38] The $V_{TH}$ shift to more negative values as the InO$_x$layer thickness increases can be explained by an increasing number of free electrons in the film that largely stem from oxygen vacancies. With thicker films, these are more stable near the channel. However, any further $V_{TH}$ shift to negative voltages past 0 V would increase the 0 V device current and so the overall power consumption of the device. Therefore, we determined the three-layer coating to be an overall optimum and did not further increase the number of layers even though the mobility would continue to increase with thickness. All of the $InO_x$-TFTs exhibit a very small clockwise hysteresis loop (positive $V_{TH}$ shift) as can be seen in Figure 5c. The appearance of clockwise hysteresis can possibly be attributed to an incomplete precursor conversion in the M-O-M network, resulting in electron traps at the dielectric-semiconductor interface during operation.[74, 75] To test whether the $V_{TH}$ of thinner films could be improved by choosing different annealing times, we also looked at 1C-TFTs for annealing times other than 60 and 300 s (Figure S15, Supporting Information), but found that in addition to other shortcomings such as larger spread of device yield, the strongly positive $V_{TH}$ of very thin films cannot be improved by different annealing times.

In the balance of achieved mobility, small non-negative $V_{TH}$, and still negligible hysteresis, we consider the triple-coated MO films (3C-TFTs) as an optimum semiconductor layer structure.

3C-TFTs in which the $InO_x$ layers were not just annealed for 60 or 300 s at 300 °C but for additional values up to 1200 s were also studied, and their electrical characterization is presented in Figure 5e. All the 3C-TFTs exhibited good electrical characteristics with small hysteresis. The discussion of the corresponding electrical performance details is given in Table S5 (Supporting Information).

*2.4.2 Heterojunction IZI-TFTs*

According to previous works on heterojunction TFTs, $InO_x$ is the best choice as the primary (1st-coated) layer in contact with the dielectric interface to maximize the free charge carrier concentration in the heterojunction devices.[76, 77] Therefore, $InO_x$-$ZnO_x$-$InO_x$ (IZI) heterojunction films were produced while keeping the film coating and annealing parameters constant for all layers. **Figure** 6a shows the results for heterojunction TFTs wherein the $ZnO_x$ layer is sandwiched between two $InO_x$ films.

The device transfer characteristics of 3C and IZI heterojunction TFTs are shown in Figure 6b while **Table** 1 lists the corresponding electrical parameters. The mobility distribution of 30 IZI-TFTs are summarized in Figure 6c. The *I–V* curves of IZI-TFTs with different annealing times are shown in Figure 6d. It is obvious that the devices annealed with long annealing treatment have higher mobility and more negative $V_{TH}$ (Figure 6e) compared to 3C-TFTs.

**Figure 6.** a) Schematic of IZI-TFTs. b) Transfer characteristics of 3C-TFTs and IZI-TFTs. c) Distribution of mobility in 30 IZI-TFTs. d) *I–V* curves of IZI-TFTs on Si/SiO$_2$ substrates with different annealing time. e) 3D plot of mobility-annealing time relationship for MO-TFTs on 300-nm Si/SiO$_2$ substrates. f) Energy band diagram of IZI heterostructure stacks.

**Table 1**. Extracted representative electrical parameters of TCIO-TFTs with different InO$_x$ annealed for different time.

| Annealing time [s] | $\mu_e$ [cm$^2$ V$^{-1}$ s$^{-1}$] | $V_{TH}$ [V] | $V_{ON}$ [V] | $I_{on}/I_{off}$ | S.S. [V·dec$^{-1}$] |
|---|---|---|---|---|---|
| 5 | – | – | – | – | – |
| 10 | 3.6 | 14.8 | −0.6 | 7.1 × 10$^6$ | 1.7 |
| 30 | 7.9 | 12.5 | −1.4 | 6.2 × 10$^6$ | 1.6 |
| 60 | 13.8 | 6.2 | −3.2 | 4.1 × 10$^6$ | 2.4 |
| 120 | 17.9 | 4.6 | −4.8 | 3.7 × 10$^6$ | 2.2 |
| 300 | 28.4 | 1.2 | −5.7 | 2.8 × 10$^6$ | 2.3 |
| 600 | 34.5 | −1.8 | −8.3 | 2.2 × 10$^6$ | 2.6 |
| 1200 | 38.2 | −3.3 | −10.3 | 1.0 × 10$^6$ | 2.8 |

Figure 6f shows the band energy diagram of the heterojunction IZI-TFT as constructed from optical band gap measurements and the UPS data under the assumption of Fermi level alignment ($E_{fn}$). The $E_{VBM}$ values of the InO$_x$ and ZnO$_x$ are extracted from XPS results of InO$_x$ and ZnO$_x$ films annealed at 300 °C for 300 s, respectively. The $E_G$ value is determined to be 3.64 eV (InO$_x$) and 3.48 (ZnO$_x$) from the optical band gap by UV–vis spectroscopy measurements. Based on the resulting energetic band offsets an energy barrier energy of 0.18 eV forms at the interfaces of $ZnO_x/InO_x$ (left) and $InO_x/ZnO_x$ (right). At such a heterojunction, charge transfer can occur from splitting of thermal and optical band-to-band excitation states, and as a result lead to concentrations of free electrons in the $InO_x$ and free holes in the $ZnO_x$ on either side of the junction.[78] This process happens to some extend in the dark and at much higher densities under illumination. From the point of view of the transistor operation, these free charges are in addition to the capacitance-limited and gate-bias controlled electron density in the channel at the $InO_x$/dielectric interface.[79] During gated operation, these extra carriers can quench existing trap states that would otherwise impede transistor operation, resulting in an improvement of mobility and *SS* at higher concentrations of these extra carriers such as when the heterojunction devices are exposed to light, $V_{TH}$ shifts can be observed (see below). A similar phenomenon of heterojunction devices with enhanced properties has been reported using the terminology 2DEG TFTs.[68, 77]

*2.4.3 MO-TFTs with $AlO_x$ as the dielectric layer*

Considering the excellent electrical performance achieved for the 3C and IZI films as semiconductor channel on $Si/SiO_2$ substrates, we also explored the use of the high-k material $AlO_x$ as device dielectric layer to achieve higher electrical performances, high stability, and low-voltage operation. In various literature reports, toxic organic solvents and complicated preparation process were usually involved in solution-processed synthesis of $AlO_x$ films[80, 81] whereas here we chose to continue to use the green solvent water.

Fabrication of $AlO_x$ films followed similar procedures as for the other MO films. 0.2 m$Al(NO_3)_3$ precursor solution were prepared as aqueous solutions. A single layer of the precursor solution was shear-coated onto the Si substrate and annealed at 300 °C for 60 s in air, resulting in a 4.6 nm film as verified via TEM. (Table S4, Supporting Information). The O1 s XPS peaks of $AlO_x$ film can be de-convoluted into two components: one located at 531.6 ± 0.2 eV, corresponding to Al–OH bonds, and the other peak located at 532.7 ± 0.2 eV, corresponding to Al–O–Al bonds (Figure S10, Supporting Information). The area ratio of Al–OH to Al–O–Al decreased with annealing time and finally saturates at 85:15 for annealing times longer than 60 s (Table S3, Supporting Information). Therefore, from economic and ecological points of view, an annealing time of 60 s at 300 °C should be sufficient to produce high quality AlO$_x$ dielectrics.

Five times-coated AlO$_x$ (5C-AlO$_x$) films with a minimum thickness of ≈20 nm and annealed at 300 °C for 60 s were then used to measure the capacitance and subsequently produce high-performance, low-voltage MO-TFTs. Capacitance–frequency (C–f) properties of capacitors with the AlO$_x$ dielectric were measured in the 20–10$^6$ Hz frequency range under the bias voltages of −1, 0, and +1 V. Due to the low processing temperature and short annealing time, the AlO$_x$ films lead to a large mobile H$^+$ content inside, which show a capacitance dispersion at low frequencies. These AlO$_x$ films exhibited excellent dielectric characteristics, as indicated by the very low leakage current density of $1.3 \times 10^{-8}$ A cm$^{-2}$, and high areal capacitance of 208 nF cm$^{-2}$ from 10$^3$ to 10$^5$ Hz with an Al electrode with a diameter of 1.0 mm. From the measured capacitance, the calculated dielectric constant of AlO$_x$ was 6.3. This result is consistent with the general observation of AlO$_x$ films with sol-gel solution process for identical temperature.[82]

Based on these measurements, we took the average value of capacitance in the range of 10$^3$ to 10$^5$ Hz when calculating mobility of AlO$_x$ based TFTs (Figure S11, Supporting Information). Leakage current density versus voltage graphs of AlO$_x$ films with different layers and different annealing treatment times are shown in Figure S11 (Supporting Information). We also fabricated capacitors of various electrode shapes and verified that the capacitance did scale with the electrode area and with the inverse of the dielectric film thickness as expected. This confirmed the spatial uniformity of the AlO$_x$ films. Furthermore, the fabrication method was found to be highly reliable in terms of film quality and thickness obtained for any given annealing treatment (Figure S12, Supporting Information). Even when extended annealing times of 1200 s were used, no Bragg diffraction features were observed in the corresponding GIWAXS images of these films, from which we infer that the AlO$_x$ films remain amorphous for short annealing times (10 to 1200 s; Figures S6 and S8, Supporting Information).

In **Figure** 7a, the IZI films with single coated AlO$_x$ dielectric films annealed at 300 °C for 60 s on quartz glass substrate also show a high degree of optical transparency in the visible light range. For the MO TFTs, 5C-AlO$_x$ films with each film annealed at 300 °C for 60 s were utilized as gate dielectric layers on Si substrates. 3C-InO$_x$ and IZI films served as the channel layer, with each MO semiconductor film annealed at 300 °C for 60 s. All the present MO-TFTs based on AlO$_x$ dielectric show typical transistor response with low operating voltage, $V_G = -1$ to 3 V for $V_{DS} = 2$ V. Note, all calculations are based on the sweeping curves when the devices exhibit acceptable hysteresis with voltage shift <0.5 V. As shown in the representative transfer and output plots in Figure 7c and Figure S13 (Supporting Information), 3C-TFTs exhibit impressive device characteristics with mobilities $\mu_e$ of ≈6.4 cm$^2$ V$^{-1}$ s$^{-1}$, $I_{on/off}$ ratio of $2.3 \times 10^3$, a threshold voltage of 1.1 V, and a subthreshold swing of 0.30 V dec$^{-1}$. The fabricated IZI-TFTs exhibited a $\mu_e$ of 13.2 cm$^2$ V$^{-1}$ s$^{-1}$, including a $V_{TH}$ of 0.6 V, a high $I_{on}/I_{off}$ of $1.7 \times 10^6$, and a $SS$ of 0.2 V dec$^{-1}$. In agreement with results on standard Si/SiO$_2$ substrates, longer annealing durations (e.g., 300 °C for 1200 s) enhanced the electron mobility. **Table** 2 summarizes the existing literature values including the respective annealing temperature and duration for InO$_x$ TFTs with AlO$_x$ dielectric layers. Compared with these values, the 3C-TFTs with AlO$_x$ dielectric and only rapid annealing (at 300 °C for 60 s) can achieve remarkable mobility.

**Figure 7.** a) Transmittance of IZI film with 1C-AlO$_x$ dielectric film on quartz substrate (Inset images show the schematic structure and optical photo of samples). b) Transfer characteristics of the 3C-TFTs and IZI-TFTs based on InO$_x$ and ZnO$_x$ films on AlO$_x$/Si substrates. c) Schematic of the flexible TFTs and photograph on the polyimide substrate.

d) Transfer characteristics of the flexible TFTs. The MO semiconductor films and AlO$_x$ dielectric layers were annealed at 300 °C for 60 s.

**Table 2**. Several results of solution-based InOx TFTs characters annealed at various temperatures and times.

| TFT Structure | Dielectric | *Solvent* | *Temperature* [°C] | Time [min] | $\mu_e$ [cm$^2$ V$^{-1}$ s$^{-1}$] | $I_{on}/I_{off}$ | *Reference* |
|---|---|---|---|---|---|---|---|
| InO$_x$ | AlO$_x$ | water | 250 | 20 | 11.5 | $1.4 \times 10^3$ | [82] |
| InO$_x$ | AlO$_x$ | water | 300 | 360 | 11.9 | $2.5 \times 10^3$ | [83] |
| InO$_x$ | AlO$_x$ | water | 300 | 360 | 15.8 | $1.0 \times 10^6$ | [84] |
| InOx | AlOx | 2-ME[a)] | 300 | 30 | 3.5 | $1.4 \times 10^5$ | [84] |
| InO$_x$ | AlO$_x$ | 2-ME | 200 | 30 | 3.0 | – | [85] |
| InO$_x$ | AlO$_x$ | 2-ME | 300 | 30 | 14.5 | – | [86] |
| InO$_x$ | AlO$_x$ | 2-ME | 350 | 30 | 0.4 | – | [87] |
| 3C-TFTs | AlO$_x$ | water | 300 | 1 | 6.4 | $2.3 \times 10^3$ | * |
| IZI-TFTs | AlO$_x$ | water | 300 | 1 | 13.2 | $2.5 \times 10^3$ | * |

[a)] 2-Methoxyethanol. * Result from this work

To demonstrate the applicability of the heterojunction films fabricated from aqueous solution process at low temperatures to flexible substrates, we also fabricated MO-TFTs with AlO$_x$ dielectric films on polyimide (PI). The BGTC device structure and an image of the as-fabricated film is illustrated in Figure 7c. Thermally evaporated 50 nm Al electrodes were defined on a PI substrate as the bottom gate electrode, and five times-coated, solution-processed AlO$_x$ served as the gate dielectric layer. The triple-coated InO$_x$ films were deposited and the flexible devices were completed by evaporation of Al source/drain electrode on the top ($W/L$ = 1000 µm/100 µm). 5 flexible TFTs were evaluated for the statistical data in this section. Clearly, it reveals that IZI-TFTs exhibit larger mobility than 3C-TFTs with an average value of 10.3 and 4.2 cm$^2$ V$^{−1}$ s$^{−1}$, respectively (Figure 7d). Therefore, although the electronic performance slightly degrades as the substrate changed from Si wafer to PI film, the IZI-TFTs with the AlO$_x$ dielectric layer still exhibit a higher mobility than the 3C-TFTs.

The TFTs using the AlO$_x$ dielectric exhibited a small but visible degree of anticlockwise hysteresis with a positive $V_{TH}$ shift. According to other reports, this phenomenon reflects a non-constant gate capacitance when the devices are measured with a gate potential at low frequencies.[75, 88] The counterclockwise hysteresis could possibly be related to hydroxyl groups in the AlO$_x$ dielectric layer.[89] In this work, the AlO$_x$ films were annealed at 300 °C for only 60 s, which likely leaves a small but non-negligible number of hydroxyl groups in the films. These hydroxyl groups could then act as possible electron trap states in AlO$_x$ leading to the observed counterclockwise hysteresis.

The mobility values of the devices on PI film (Figure 7d) are lower than the devices fabricated on the Si substrate (Figure 7c), likely due to the higher substrate roughness of the PI film compared to the SiO$_2$ wafer surface that is inherited by the subsequently deposited layers (including the Al gate and thus the AlO$_x$ dielectric film). The AlO$_x$ dielectric films on Al/PI substrate surface exhibited a $\sigma_{RMS}$ of 1.1 ± 0.3 nm, while the AlO$_x$ dielectric films on Si substrates exhibited a $\sigma_{RMS}$ of 0.4 ± 0.2 nm. It is well known that charge carriers are more easily scattered when the dielectric/semiconductor interface is rough.[90] Another possibility is that the Al gate could to a small degree react with water when the solution processed AlO$_x$ dielectric layer is being annealed, with a possible reaction product being AlOOH.[91]

*2.4.4 Device Stability*

Since long-term stability is another important quality of devices, positive gate-bias stress (PBS) testing was also performed by applying a positive potential of + 60 V with a stress time of 5000 s to the gate terminal at RT in air. The transfer characteristics were measured at a fixed $V_{DS}$ = +80 V to investigate variation in the performance of the TFTs under stress without encapsulation. **Figure** 8a shows the $V_{TH}$ shift of 12.6 and 11.5 V in 3C-TFT and IZI-TFT, respectively, after 5000 s of PBS. Here, PBS induces a positive $V_{TH}$ shift. The observed $V_{TH}$ shift can be attributed to the adsorption of oxygen molecules from ambient air, which combine with electrons in the metal oxide film therefore creating more traps in the channel.[92, 93] Both the $I_{on}/I_{off}$ and SS in 3C-TFT and IZI-TFT slightly increase due to the gate-bias stress induced off-current decrease.[94] The slightly increased SS value implies a generation of fixed space charges at the dielectric interface.[68] This trend is similar to that of previously reported devices.[41, 52, 95] In term of bias stability, the solution processed MO-TFTs exhibit relatively good stability under gate-bias stress condition with an acceptable degree of $V_{TH}$ shift.

Figure 8. The time dependence of transfer curves measured at a fixed $V_{DS}$ = +80 V and mobility distribution from 30 devices after a) PBS and b) NBIS testing for the 3C-TFT and IZI-TFT. Mobility distribution of 30 devices of c) 3C-TFTs and d) IZI-TFTs aged for two months.

However, it has been pointed out in literature that for any technology to be industrially relevant, is also important to study the combination of light illumination and gate-bias.[96] Other groups have reported that light in the green range can cause instability problems during the characterization of MO-TFTs.[40, 97] Here, to investigate the stability of the fabricated TFTs under light illumination, 3C-TFTs and IZI-TFTs measurements were performed under exposure to a 565 nm green light with a power of 5 µW cm$^{-2}$ in a dark room. We applied a negative gate bias of −60 V for 500 s. Figure 8b shows the changes in $I_{DS}$-$V_G$ transfer curves of all TFTs before and after negative light bias stress (NLBS) respectively for different time at a fixed $V_{DS}$ = +80 V. Time evolution with light illumination for 5, 10, 15, 30, 45, and 60 s on the MO-TFTs behavior were also characterized. There are obvious $V_{TH}$ shifts for 3C-TFTs (3.4 V) and IZI-TFTs (4.7 V) toward the negative direction. In view of the n-type characteristics of

MO, this shift is typically referred to the trapping of photo excited holes in the MO layer, which is capable to modulate the capacitance and mobility of the MO channel.

In the NLBS test, the combination of green light illumination and gate-bias typically caused the negative $V_{TH}$ shift and increased current in all the devices. It has been proposed that such behavior originates from photo-induced hole trapping at trapping centers in the gate dielectric layer.[98] Photo-induced electrons and holes are generated in the semiconductor layer through transformation of electron-hole pairs. The so-generated holes are initially trapped in the deep-level states in the channel and subsequently injected into defect states of the gate dielectric. The generated electrons become carriers in the MO-TFTs which increases the current during the measurement. The injected holes in the gate dielectric could play a role in positive charges, causing the $V_{TH}$ shift. Since $InO_x$ and $ZnO_x$ have significant numbers of deep-level states caused by oxygen vacancy defects, all the devices shown a negative $V_{TH}$ shift.[99] In contrast, IZI-TFTs exhibit better NLBS stability than 3C-TFTs. We presume this phenomenon is due to $ZnO_x$ having a smaller proportion of oxygen vacancies than $InO_x$ (Tables S1 and S2, Supporting Information), which means fewer deep-level states exist in IZI films. Therefore, it is reasonable to see that the IZI-TFTs exhibit better NLBS stability.

From the bias stress test result, we found that the heterojunction IZI-TFTs not only have improved electrical properties of $\mu_e$ and *SS*, but also provide advanced gate bias stress stability and better devices against light sensitivity. Obtaining 3C-TFTs and IZI-TFTs with air-stability against oxygen and atmospheric moisture is another important target. The mobility distribution of 30 samples of 3C-TFTs and IZI-TFTs that were simply kept unsealed on a shelf in ambient conditions are shown Figure 8c,d. Most of devices still maintain high mobility after 2 months.

**3 Conclusions**

In this study, we demonstrate solution-processed, high performance MO-TFTs, fabricated by blade coating with thermal annealing using water as the solvent. We show the conversion from the aqueous precursors to nanometer-thin metal oxide thin films at low temperatures of 300 °C for different annealing times with the ellipsometry and TEM techniques. Our combined XPS and UPS studies show that the dehydroxylation and dehydration of the hydrogen-related species and the formation of oxygen vacancies occur within the time frame of 10–300 s. UPS and UV–vis data suggest a raise in both valance band maximum and Fermi level occurs with longer annealing treatment.

Importantly, we find that the annealing time of MO film can be significantly reduced compared to often reported values such as ten of minutes or even several hours. Here, MO-TFTs are obtained with a blade coating process after only a few minutes of annealing. The triple-coated $InO_x$-TFTs (3C-TFTs) annealed at 300 °C for 60 s on a 300 nm $SiO_2$ gate dielectric exhibits a high field effect mobility of 10.7 $cm^2\ V^{-1}\ s^{-1}$.

By engineering a $ZnO_x$ layer between $InO_x$ layers, we obtained sandwiched IZI-TFTs with enhanced electrical (≈13.8 $cm^2\ V^{-1}\ s^{-1}$) and outstanding optical properties compared to the 3C-TFTs by the rapid annealing approach (60 s). We also obtained high-performance 3C-TFTs (over 18 $cm^2\ V^{-1}\ s^{-1}$) and IZI-TFTs (over 38 $cm^2\ V^{-1}\ s^{-1}$) by long-time annealing treatment at 300 °C for 1200 s. These results demonstrate that the strategy of homojunction semiconductor layer of IZI films enhance not only the TFT electron mobility, but also the positive bias and light bias stability. Furthermore, flexible IZI-TFTs with semiconductor and $AlO_x$ dielectric films by rapid annealing (at 300 °C for 60 s) can achieve a remarkable mobility of over 10 $cm^2\ V^{-1}\ s^{-1}$ at low operating voltage.

In summary, this study paves a way toward not only a simple, sustainable and eco-friendly process on developing high performance MO films, but also a feasible path to flexible devices for advanced low-power electronics.

**4 Experimental Section**

*Preparation of Sol-Gel Precursor Solution*

All reagents were purchased from Sigma–Aldrich. Indium nitrate (In(NO$_3$)$_3\cdot x$H$_2$O, 99.99%), zinc nitrate (Zn(NO$_3$)$_3\cdot$6H$_2$O, >99.0%), and aluminum nitrate nonahydrate (Al(NO$_3$)$_3\cdot$9H$_2$O, 99.997%) were dissolved in deionized water to achieve the precursor solutions (0.1 m of In(NO$_3$)$_3$ and Zn(NO$_3$)$_3$, 0.2 M of Al(NO$_3$)$_3$). All the solutions were stirred at 500 rpm for 6 h and aged for 6 h at room temperature under ambient conditions and filtered using a 0.2 µm PTFE filter prior to film fabrication.

*Oxide Film Characterization*

Topography and roughness of semiconductor films images were obtained in tapping mode in an area of 5 µm$^2$ with a Flex Axiom from Nanosurf atomic force microscopy (AFM) and TAP-190Al-G tips from Budget Sensors. Photoemission spectroscopy measurements are performed with a Thermofisher Escalab 250Xi system.

UV–vis spectroscopy measurements were performed with a Cary 5000 (Agilent, USA) spectrometer. Photoemission spectroscopy measurements were performed with a Thermofisher Escalab 250Xi system. The samples were measured in the dark in ultrahigh vacuum (10$^{-10}$ bar). UPS measurements are performed with a He I lamp (21.22 eV), an applied bias of −5 V and a pass energy of 2 eV.

Ellipsometry measurements were done with an EP4 imaging, spectrocsopic ellipsometer (Accurion, Germany). For single layers on silicon substrates (native oxide) prepared with an identical fabrication protocol but different annealing times, the ellipsometric parameters $\Delta$ and $\Psi$ have been measured within a wavelength range of 360–1000 nm and at two incidence angles (60° and 70°). An optical model including three Gaussian oscillators was applied to determine the optical constants and layer thicknesses.

TEM was conducted using a JEOL JEM F200 transmission electron microscope operated at 200 kV acceleration voltage. (Energy Dispersive spectroscopy). EDS analysis was performed using a dual 100-mm$^2$ windowless silicon drift detector. The EDS spectra were de-noised with principal component analyses using three components.

*Grazing-Incidence Wide-Angle X-Ray Scattering (GIWAXS)*

The GIWAXS measurements were carried out at the ID10 beamline at the European Synchrotron Radiation Facility (ESRF). The beam energy was 10 keV and the incidence angle was 0.14. The images were recorded with a Dectris Pilatus 300K area detector, which was placed ≈16 cm behind the sample. The exposure time was between 1 and 5 s. The measurements were calibrated using a LaB$_6$ scattering standard. The analysis was performed with the WxDiff software.

*Devices Fabrication and Measurement*

Heavily doped silicon (Si$^{++}$) wafers with and without a thermally grown 300 nm SiO$_2$ layer were used as substrates for the devices. For the transparent substrate, 1.0 inch × 1.0 inch quartz substrates were used in this work. All the substrates were cleaned by sonication in acetone and isopropanol bath lasting for 10 min. Post further drying using N$_2$ gas, the substrates were exposed to UV plasma for ≈10 min before blade coating to get clean and low contact angle substrates.[100]

For the MO films on Si/SiO$_2$ substrate fabrication, the MO precursor solution was coated on Si$^{++}$ wafer with 300 nm SiO$_2$ layer substrates. Self-assembled monolayers formed by octadecyltrimethoxysilane (ODTMS) were grown on Si/SiO$_2$ substrate to get the blade with a good hydrophobicity, the blade shows the water contact angle value of ≈113°.[101] The blade angle was set at 8° with a gap of 30 µm between the edge of the blade and the substrate.[102] 2–5 µL of precursor solution for 1.5 inch × 0.5 inch Si/SiO$_2$ substrate was injected at the interface between the substrate and the blade according to the materials.[103] During coating, the substrate temperature was 80 °C and the blade move speed was 5000 µm s$^{-1}$. The solution shearing parameters were optimized in an initial assessment of the resulting TFT device performance and then kept fixed throughout this study (Figure S14, Supporting Information). After coating, the samples were transferred on to a hot plate set at the target temperature of 300 °C with varied time (10–1200 s). This process was repeated different times to achieve the desired semiconductor film thickness. Finally, aluminum source and drain electrodes with thickness of 50 nm) were deposited by thermal evaporation using shadow masks under a very high vacuum of $1 \times 10^{-7}$ mbar with a deposition velocity of 0.5 Å s$^{-1}$. For the MO-TFTs based on AlO$_x$ dielectric films, the AlO$_x$ precursor solution was coated on bare Si$^{++}$ wafers and annealed at 300 °C to form AlO$_x$ dielectric layer.

The flexible devices were fabricated by evaporating 50 nm of Al on polyimide film as gate electrode followed by blade coated AlO$_x$ as a dielectric layer. This process was repeated different times to obtain the desired dielectric layers. For capacitors, 50 nm Al was evaporated through a shadow mask to get three different size electrodes. with diameter circles of 1.0, 0.5, and 0.25 mm, respectively.

All devices were measured utilizing a Keysight B1500 Semiconductor Analyzer in a dark room. The electron field-effect mobility ($\mu_e$) was calculated using the equation $\mu_e = 2 \times LWC_i \times (\partial |I_{DS}|\partial V_{GS})^2$ by averaging the slopes in the range of source and drain voltage. Here, $I_{DS}$ is the drain current, $C_i$ is the capacitance per unit area of the dielectric, $W$ is the channel width, $L$ is the channel length, and $V_{GS}$ is the gate voltage. The $V_{TH}$ was obtained from the intercept of the linear fit of $I_{DS}^{1/2}$ versus $V_{GS}$ curve and the $x$-axis. In this report, the channel width and length for all TFTs were 1000 and 100 µm, respectively. The average mobility was calculated from all devices, only excluding those that had large-scale optically visible film defects (e.g., edge effect inhomogeneities at the beginning/end of the blade coating area). The coating, annealing, and morphology measurements were performed in ambient atmosphere with relative humidity ≈30%.

**Acknowledgements**

T.T. is thankful for the financial support from the China Scholarship Council Ph.D. Program (no. 201706310174). K.H. acknowledges the funding of the European Social

Fund (Project OrgNanoMorph, proposal no. 100382168). F.T. and S.M. acknowledge financial support from the German Research Foundation (DFG, MA 3342/6-1). J.P. and Y.K. acknowledge the financial support from German Research Foundation (DFG, KR 4364/4). The authors would like to acknowledge support by the German Excellence Initiative via the Cluster of Excellence EXC 1056 "Center for Advancing Electronics Dresden" (CFAED). The GIWAXS experiments were performed on beamline ID10 at the European Synchrotron Radiation Facility (ESRF), Grenoble, France. The authors are grateful to Maciej Jankowski and Oleg Konovalov at the ESRF for providing assistance in using beamline ID10. This project has received funding from the European Research Council (ERC) under the European Union's Horizon 2020 research and innovation programme (ERC Grant Agreement No. 714067, ENERGYMAPS). The authors acknowledge the use of the facilities in the Dresden Center for Nanoanalysis (DCN) at the Technische Universität Dresden.

Open access funding enabled and organized by Projekt DEAL.

**Conflict of Interest**

The authors declare no conflict of interest.

**Author Contributions**

T. Tang designed the experiment and fabricated all the samples. J. Kreß performed the XPS and UPS measurements and analysis under the supervision of Y. Vaynzof. C. Hänisch performed the ellipsometry measurements under the supervision of S. Reineke. P. Dacha, J.Perez, F. Talnack and K. Haase contributed the experimental design and data analysis. A. Tahn and D. Pohl performed the film characterization. T. Tang and P. Dacha wrote the manuscript. All authors approved to the final version of the manuscript.